\documentclass[10pt]{article}
\usepackage{amsfonts,amsmath,amssymb}
\usepackage[T1,T5]{fontenc}
\usepackage[Gray,squaren,thinqspace,thinspace]{SIunits}
\usepackage[dvips]{feynmp}
\usepackage{epsfig}
\usepackage{dsfont}
\usepackage{rotating}
\usepackage{float,braket}
\usepackage{subfigure}
\usepackage{amsmath}
\usepackage{amsfonts}
\usepackage{amssymb}
\usepackage{graphicx,wrapfig,color}

\newcommand{\MSbar}{\overline{\mbox{MS}}}

\newcommand{\p}{\partial}

\DeclareMathOperator{\tr}{tr}

\usepackage{color}
\definecolor{darkgreen}{rgb}{0,0.35,0}

\definecolor{Rood}{rgb}{1, 0, 0}

\begin{document}

\title{\noindent \textbf{Effect of the Gribov horizon on the Polyakov loop
and vice versa}}
\author{\textbf{F.~E.~Canfora$^{a}$}\thanks{%
fcanforat@gmail.com}\,\,, \textbf{D.~Dudal}$^{b,c}$\thanks{%
david.dudal@kuleuven-kulak.be}\,\,, \textbf{I.~F.~Justo}$^{c,d}$\thanks{%
igorfjusto@gmail.com}\,\,, \\\textbf{P.~Pais}$^{a,e}$\thanks{%
pais@cecs.cl}\,\,, \textbf{L.~Rosa}$^{f,g}$\thanks{%
rosa@na.infn.it}\,\,,
\textbf{D.~Vercauteren}$^{h}$\thanks{vercauterendavid@dtu.edu.vn}\\
{\small \textnormal{$^{a}$ Centro de Estudios Cient\'{\i}ficos (CECS),
Casilla 1469, Valdivia, Chile}}\\
{\small \textnormal{$^{b}$ KU Leuven Campus Kortrijk - KULAK, Department of
Physics, }} \\
{\small \textnormal{\phantom{$^{b}$} Etienne Sabbelaan 53, 8500 Kortrijk, Belgium}}\\
{\small \textnormal{$^{c}$ Ghent University, Department of Physics and
Astronomy}}\\
{\small \textnormal{\phantom{$^{c}$}  Krijgslaan 281-S9, 9000 Gent, Belgium}}\\
{\small \textnormal{$^{d}$ Departamento de F\'{\i}sica Te\'{o}rica,
Instituto de F\'{\i}sica, UERJ - Universidade do Estado do Rio de Janeiro,}}%
\\
{\small \textnormal{\phantom{$^{d}$} Rua S\~{a}o Francisco Xavier 524,
20550-013 Maracan\~{a}, Rio de Janeiro, Brazil }}\\
{\small \textnormal{$^{e}$ Physique Th\'{e}orique et Math\'{e}matique, Universit\'{e} Libre de Bruxelles}} \\
{\small \textnormal{\phantom{$^{e}$} and International Solvay Institutes, Campus Plaine C.P.~231, B-1050 Bruxelles, Belgium}}\\
{\small \textnormal{$^{f}$ Dipartimento di Fisica, Universit\'{a} di Napoli
Federico II, Complesso Universitario di Monte S.~Angelo,}}\\
{\small \textnormal{\phantom{$^{f}$} Via Cintia Edificio 6, 80126 Napoli,
Italia}}\\
{\small \textnormal{$^{g}$ INFN, Sezione di Napoli, Complesso Universitario
di Monte S.~Angelo,}}\\
{\small \textnormal{\phantom{$^{g}$} Via Cintia Edificio 6, 80126 Napoli, Italia}}\\
{\small \textnormal{$^{h}$ Duy T\^an University, Institute of Research and
Development,}}\\{\small \textnormal{\phantom{$^{h}$} P809, K7/25 Quang Trung, {\fontencoding{T5}\selectfont H\h ai
Ch\^au, \DJ \`a N\~\abreve ng}, Vietnam}} }
\date{}
\maketitle

\begin{abstract}
We consider finite temperature $SU(2)$ gauge theory in the continuum formulation, which necessitates the choice of a gauge fixing. Choosing the Landau gauge, the existing gauge copies are taken into account by means of the Gribov--Zwanziger (GZ) quantization scheme, which entails
the introduction of a dynamical mass scale (Gribov mass) directly influencing the Green functions of the theory. Here, we determine simultaneously the Polyakov loop (vacuum expectation value) and Gribov mass in terms of temperature, by minimizing the vacuum energy w.r.t. the Polyakov loop parameter and solving the Gribov gap equation. Inspired by the Casimir energy-style of computation, we illustrate the usage of Zeta function regularization in finite temperature calculations. Our main result is that the Gribov mass directly feels the deconfinement transition, visible from a cusp occurring at the same temperature where the Polyakov loop becomes nonzero. In this exploratory work we mainly restrict ourselves to the original Gribov--Zwanziger quantization procedure in order to illustrate the approach and the potential direct link between the vacuum structure of the theory (dynamical mass scales) and (de)confinement. We  also present a first look at the critical temperature obtained from the Refined Gribov--Zwanziger approach. Finally, a particular problem for the pressure at low temperatures is reported.
\end{abstract}

\section{Introduction}
Within $SU(N)$ Yang-Mills gauge theories, it is well accepted that
the asymptotic particle spectrum does not contain the elementary excitations
of quarks and gluons. These color charged objects are confined into color
neutral bound states: this is the so-called color confinement phenomenon. It
is widely believed that confinement arises due to non-perturbative infrared
effects. Many criteria for confinement have been proposed (see the nice
pedagogical introduction \cite{Greensite:2011zz}). A very natural observation is that gluons (due to the fact that they are
not observed experimentally) should not belong to the physical spectrum in a
confining theory. On the other hand, the perturbative gluon propagator
satisfies the criterion to belong to the physical spectrum (namely, it has a
K{\"{a}}ll{\'{e}}n-Lehmann spectral representation with positive spectral
density). Hence, non-perturbative effects must dress the perturbative
propagator in such a way that the positivity conditions are violated, such that it does not belong to the physical spectrum anymore. A well-known
criterion is related to the fact that the Polyakov loop \cite{Polyakov:1978vu} is
an order parameter for the confinement/deconfinement phase transition via its connection to the free energy of a (very heavy) quark. The importance to clarify the interplay between these two
different points of view (nonperturbative Green function's behaviour vs.~Polyakov loop) can be understood by observing that while there
are, in principle, infinitely many different ways to write down a gluon
propagator which violates the positivity conditions, it is very likely that
only few of these ways turns out to be compatible with the Polyakov
criterion.

One of the most fascinating non-perturbative infrared effects is related to the
appearance of Gribov copies \cite{Gribov:1977wm} which represent an
intrinsic overcounting of the gauge-field configurations which the
perturbative gauge-fixing procedure is unable to take care of. Soon after Gribov's seminal paper, Singer showed that any true gauge condition, as the Landau gauge\footnote{We shall work exclusively with the Landau gauge here.},
presents this obstruction \cite{Singer:1978dk} (see also \cite{Jackiw:1977ng}). The
presence of Gribov copies close to the identity induces the existence of
non-trivial zero modes of the Faddeev-Popov operator, which make the path
integral ill defined. Even when perturbation theory around the  vacuum is not
affected by Gribov copies close to the identity (in particular, when
YM-theory is defined over a flat space-time\footnote{%
In the curved case, the pattern of appearance of Gribov copies can be
considerably more complicated: see in particular \cite{P1,P2,P3,P4}.
Therefore, only the flat case will be considered here.} with trivial topology
\cite{Sobreiro:2005ec}), Gribov copies have to be taken into account when considering
more general cases (such as with toroidal boundary conditions on flat
space-time \cite{hedgehog,degenerate-systems}). Thus, in the following only the standard
boundary conditions will be considered.

The most effective method to eliminate Gribov copies, at leading order proposed by Gribov
himself, and refined later on by Zwanziger \cite{Gribov:1977wm,Zwanziger:1988jt,Zwanziger:1989mf,Zwanziger:1992qr} corresponds to restricting the path integral to the so-called Gribov region,
which is the region in the functional space of gauge potentials over which
the Faddeev-Popov operator is positive definite. The Faddeev--Popov operator is Hermitian in the Landau gauge, so it makes sense to discuss its sign. In \cite{Dell'Antonio:1989jn,Dell'Antonio:1991xt} Dell'Antonio
and Zwanziger showed that all the orbits of the theory intersect the Gribov
region, indicating that no physical information is lost when implementing
this restriction. Even though this region still contains copies which are
not close to the identity \cite{vanBaal:1991zw}, this restriction has remarkable
effects. In fact, due to the presence of a dynamical (Gribov) mass scale, the gluon propagator is suppressed while the ghost propagator is enhanced in the infrared. More general, an approach in which the gluon propagator is ``dressed'' by
non-perturbative corrections which push the gluon out of the physical
spectrum leads to propagators and glueball masses in agreement with the
lattice data \cite{Dudal:2010cd,Dudal:2013wja}. With the same approach, one can also solve
the old problem of the Casimir energy in the MIT-bag model \cite%
{Canfora:2013zna}. Moreover, the extension of the Gribov gap equation at
finite temperature provides one with a good qualitative understanding,
already within the semiclassical approximation, of the deconfinement
temperature as well as of a possible intermediate phase in which features of the confining phase coexist with features of the fully
deconfined phase in agreement with different approaches (see \cite{Canfora:2013kma} and references therein). Furthermore, within this
framework the presence of the Higgs field \cite{Capri:2012ah,Capri:2012cr} as well of a Chern-Simons term in 2+1 dimensions \cite{Canfora:2013zza} can be accounted for
as well.

For all these reasons, it makes sense to compute the vacuum
expectation value of the Polyakov loop when we eliminate the Gribov copies using the Gribov--Zwanziger (GZ) approach. Related computations are available using different techniques to cope with nonperturbative propagators at finite temperature, see e.g.~\cite{Maas:2011se,Braun:2007bx,Marhauser:2008fz,Reinhardt:2012qe,Reinhardt:2013iia,Heffner:2015zna,Reinosa:2014ooa,Reinosa:2014zta,Fischer:2009gk,Herbst:2013ufa,Bender:1996bm}. In the present paper, we will perform for the first time (to the best of the authors' knowledge) this computation, using two different techniques, to the leading one-loop approximation. In \cite{Zwanziger:2006sc,Lichtenegger:2008mh,Fukushima:2013xsa}, it was already pointed out that the Gribov--Zwanziger quantization offers an interesting way to illuminate some of the typical infrared problems for finite temperature gauge theories.

In Section 2, we provide a brief technical overview of the Gribov--Zwanziger quantization process and eventual effective action. In the following Section 3, the Polyakov loop is introduced into the GZ theory via the background field method, building on work of other people \cite{Braun:2007bx,Marhauser:2008fz,Reinosa:2014ooa}. Next, Section 4 handles the technical computation of the leading order finite temperature effective action, while in Section 5 we discuss the gap equations, leading to our estimates for both Polyakov loop and Gribov mass. The key finding is a deconfinement phase transition at the same temperature at which the Gribov mass develops a cuspy behaviour. We subsequently also discuss the pressure and energy anomaly. Due to a problem with the pressure in the GZ formalism (regions of negativity), we take a preliminary look at the situation upon invoking the more recently developed Refined Gribov--Zwanziger approach. We summarize in Section 7.

\section{A brief summary of the Gribov--Zwanziger action in Yang--Mills
theories}

Let us start by giving a short overview of the Gribov--Zwanziger framework
\cite{Gribov:1977wm,Zwanziger:1988jt,Zwanziger:1989mf,Zwanziger:1992qr}. As
already mentioned in the Introduction, the Gribov--Zwanziger action arises
from the restriction of the domain of integration in the Euclidean
functional integral to the Gribov region $\Omega $, which is
defined as the set of all gauge field configurations fulfilling the Landau
gauge, $\partial _{\mu }A_{\mu }^{a}=0$, and for which the Faddeev--Popov
operator $\mathcal{M}^{ab}=-\partial _{\mu }(\partial _{\mu }\delta
^{ab}-gf^{abc}A_{\mu }^{c})$ is strictly positive, namely
\begin{equation*}
\Omega \;=\;\{A_{\mu }^{a}\;;\;\;\partial _{\mu }A_{\mu }^{a}=0\;;\;\;%
\mathcal{M}^{ab}=-\partial _{\mu }(\partial _{\mu }\delta
^{ab}-gf^{abc}A_{\mu }^{c})\;>0\;\}\;.
\end{equation*}%
The boundary $\p\Omega$ of the region $\Omega$ is the (first) Gribov horizon.

One starts with the Faddeev--Popov action in the Landau gauge
\begin{equation}
S_{\text{FP}}\ =S_{\text{YM}}\ +S_{\text{gf}}\ \;,  \label{fp}
\end{equation}%
where $S_{\text{YM}}$ and $S_{\text{gf}}$ denote, respectively, the
Yang--Mills and the gauge-fixing terms, namely
\begin{equation}
S_{\text{YM}}\ =\frac{1}{4}\int d^{4}x\;F_{\mu \nu }^{a}F_{\mu \nu }^{a}\;,
\label{YM}
\end{equation}%
and
\begin{equation}
S_{\text{gf}}\ =\int d^{4}x\left( b^{a}\partial _{\mu }A_{\mu }^{a}+\bar{c}%
^{a}\partial _{\mu }D_{\mu }^{ab}c^{b}\right) \;,  \label{gf}
\end{equation}%
where $({\bar{c}}^{a},c^{a})$ stand for the Faddeev--Popov ghosts, $b^{a}$
is the Lagrange multiplier implementing the Landau gauge, $D_{\mu
}^{ab}=(\delta ^{ab}\partial _{\mu }-gf^{abc}A_{\mu }^{c})$ is the covariant
derivative in the adjoint representation of $SU(N)$, and $F_{\mu \nu }^{a}$
denotes the field strength:
\begin{equation}
F_{\mu \nu }^{a}=\partial _{\mu }A_{\nu }^{a}-\partial _{\nu }A_{\mu
}^{a}+gf^{abc}A_{\mu }^{b}A_{\nu }^{c}\;.  \label{fstr}
\end{equation}%
Following \cite%
{Gribov:1977wm,Zwanziger:1988jt,Zwanziger:1989mf,Zwanziger:1992qr}, the
restriction of the domain of integration in the path integral is achieved by
adding to the Faddeev--Popov action $S_{\text{FP}}$ an additional term $%
H(A)$, called the horizon term, given by the following non-local expression
\begin{equation}
H(A,\gamma )={g^{2}}\int d^{4}x\;d^{4}y\;f^{abc}A_{\mu }^{b}(x)\left[
\mathcal{M}^{-1}(\gamma )\right] ^{ad}(x,y)f^{dec}A_{\mu }^{e}(y)\;,
\label{hf1}
\end{equation}%
where $\mathcal{M}^{-1}$ stands for the inverse of the Faddeev--Popov
operator. The partition function can then be written as \cite%
{Gribov:1977wm,Zwanziger:1988jt,Zwanziger:1989mf,Zwanziger:1992qr}:
\begin{equation}
Z_{\text{GZ}}\ =\int_{\Omega }\mathcal{D}A\;\mathcal{D}c\;\mathcal{D}\bar{c}%
\;\mathcal{D}b\;e^{-S_{\text{FP}}}=\int \mathcal{D}A\;\mathcal{D}c\;\mathcal{%
D}\bar{c}\;\mathcal{D}b\;e^{-(S_{\text{FP}}+\gamma ^{4}H(A,\gamma )-V\gamma
^{4}4(N^{2}-1))}\;,  \label{zww1}
\end{equation}%
where $V$ is the Euclidean space-time volume. The parameter $\gamma $ has
the dimension of a mass and is known as the Gribov parameter. It is not a
free parameter of the theory. It is a dynamical quantity, being determined
in a self-consistent way through a gap equation called the horizon condition
\cite{Gribov:1977wm,Zwanziger:1988jt,Zwanziger:1989mf,Zwanziger:1992qr},
given by
\begin{equation}
\left\langle H(A,\gamma )\right\rangle _{\text{GZ}}\ =4V\left(
N^{2}-1\right) \;,  \label{hc1}
\end{equation}%
where the notation $\left\langle H(A,\gamma )\right\rangle _{\text{GZ}}$
means that the vacuum expectation value of the horizon function $H(A,\gamma )
$ has to be evaluated with the measure defined in Eq.\eqref{zww1}. An
equivalent all-order proof of eq.\eqref{hc1} can be given within the
original Gribov no-pole condition framework \cite{Gribov:1977wm}, by looking
at the exact ghost propagator in an external gauge field \cite{Capri:2012wx}.

Although the horizon term $H(A,\gamma )$, eq.\eqref{hf1}, is non-local, it
can be cast in local form by means of the introduction of a set of auxiliary
fields $(\bar{\omega}_{\mu }^{ab},\omega _{\mu }^{ab},\bar{\varphi}_{\mu
}^{ab},\varphi _{\mu }^{ab})$, where $(\bar{\varphi}_{\mu }^{ab},\varphi
_{\mu }^{ab})$ are a pair of Bosonic fields, while $(\bar{\omega}_{\mu
}^{ab},\omega _{\mu }^{ab})$ are anti-commuting. It is not difficult to
show that the partition function $Z_{\text{GZ}}$ in eq.\eqref{zww1} can be
rewritten as \cite{Zwanziger:1988jt,Zwanziger:1989mf,Zwanziger:1992qr}
\begin{equation}
Z_{\text{GZ}}\ =\int \mathcal{D}\Phi \;e^{-S_{\text{GZ}}[\Phi ]}\;,
\label{lzww1}
\end{equation}%
where $\Phi $ accounts for the quantizing fields, $A$, $\bar{c}$, $c$, $b$, $%
\bar{\omega}$, $\omega $, $\bar{\varphi}$, and $\varphi $, while $S_{\text{GZ}%
}[\Phi ]$ is the Yang--Mills action plus gauge fixing and Gribov--Zwanziger
terms, in its localized version,
\begin{equation}
S_{\text{GZ}}\ =S_{\text{YM}}\ +S_{\text{gf}}\ +S_{0}+S_{\gamma }\;,
\label{sgz}
\end{equation}%
with
\begin{equation}
S_{0}=\int d^{4}x\left( {\bar{\varphi}}_{\mu }^{ac}(-\partial _{\nu }D_{\nu
}^{ab})\varphi _{\mu }^{bc}-{\bar{\omega}}_{\mu }^{ac}(-\partial _{\nu
}D_{\nu }^{ab})\omega _{\mu }^{bc}+gf^{amb}(\partial _{\nu }{\bar{\omega}}%
_{\mu }^{ac})(D_{\nu }^{mp}c^{p})\varphi _{\mu }^{bc}\right) \;,  \label{s0}
\end{equation}%
and
\begin{equation}
S_{\gamma }=\;\gamma ^{2}\int d^{4}x\left( gf^{abc}A_{\mu }^{a}(\varphi
_{\mu }^{bc}+{\bar{\varphi}}_{\mu }^{bc})\right) -4\gamma ^{4}V(N^{2}-1)\;.
\label{hfl}
\end{equation}%
It can be seen from \eqref{zww1} that the horizon condition \eqref{hc1}
takes the simpler form
\begin{equation}
\frac{\partial \mathcal{E}_{v}}{\partial \gamma ^{2}}=0\;,  \label{ggap}
\end{equation}%
which is called the gap equation. The quantity $\mathcal{E}_{v}(\gamma)$ is the vacuum energy defined by
\begin{equation}
	e^{-V\mathcal{E}_{v}}=Z_\text{GZ}\;  \label{vce} \;.
\end{equation}

The local action $S_{\text{GZ}}$ in eq.\eqref{sgz} is known as the
Gribov--Zwanziger action. Remarkably, it has been shown to be renormalizable
to all orders \cite%
{Zwanziger:1988jt,Zwanziger:1989mf,Zwanziger:1992qr,Maggiore:1993wq,Dudal:2007cw,Dudal:2008sp,Dudal:2010fq,Dudal:2011gd}%
. This important property of the Gribov--Zwanziger action is a consequence
of an extenstive set of Ward identities constraining the quantum corrections in general and possible divergences in particular.
In fact, introducing the nilpotent BRST transformations
\begin{eqnarray}
sA_{\mu }^{a} &=&-D_{\mu }^{ab}c^{b}\;,  \notag  \label{brst1} \\
sc^{a} &=&\frac{1}{2}gf^{abc}c^{b}c^{c}\;,  \notag \\
s{\bar{c}}^{a} &=&b^{a}\;,\qquad \;\;sb^{a}=0\;,  \notag \\
s{\bar{\omega}}_{\mu }^{ab} &=&{\bar{\varphi}}_{\mu }^{ab}\;,\qquad s{\bar{%
\varphi}}_{\mu }^{ab}=0\;,  \notag \\
s{\varphi }_{\mu }^{ab} &=&{\omega }_{\mu }^{ab}\;,\qquad s{\omega }_{\mu
}^{ab}=0\;,
\end{eqnarray}%
it can immediately be checked that the Gribov--Zwanziger action exhibits a soft
breaking of the BRST symmetry, as summarized by the equation
\begin{equation}
sS_{\text{GZ}}\ =\gamma ^{2}\Delta \;,  \label{brstbr}
\end{equation}%
where
\begin{equation}
\Delta =\int d^{4}x\left( -gf^{abc}(D_{\mu }^{am}c^{m})(\varphi _{\mu }^{bc}+%
{\bar{\varphi}}_{\mu }^{bc})+gf^{abc}A_{\mu }^{a}\omega _{\mu }^{bc}\right)
\;.  \label{brstb1}
\end{equation}%
Notice that the breaking term $\Delta$ is of dimension two in the fields.
As such, it is a soft breaking and the ultraviolet divergences can be controlled at the quantum level. The properties of the soft breaking of the BRST symmetry of the Gribov--Zwanziger theory and its relation with
confinement have been object of intensive investigation in recent years, see
\cite%
{Baulieu:2008fy,Dudal:2009xh,Sorella:2009vt,Sorella:2010it,Capri:2010hb,Dudal:2012sb,Reshetnyak:2013bga,Cucchieri:2014via,Capri:2014bsa,Schaden:2014bea}. Here, it suffices to mention that the broken identity \eqref{brstbr} is
connected with the restriction to the Gribov region $\Omega $. However, a set of BRST invariant composite
operators whose correlation functions exhibit the K{\"{a}}ll{\'{e}}n-Lehmann
spectral representation with positive spectral densities can be consistently
introduced \cite{Baulieu:2009ha}. These correlation functions can be
employed to obtain mass estimates on the spectrum of the glueballs \cite{Dudal:2010cd,Dudal:2013wja}.

Let us conclude this brief review of the Gribov--Zwanziger action by
noticing that the terms $S_\text{gf}$ and $S_{0}$ in expression \eqref{sgz}
can be rewritten in the form of a pure BRST variation, i.e.
\begin{equation}
S_\text{gf} + S_{0} = s \int d^4x \left( {\bar c}^a \partial_\mu A^a_\mu + {%
\bar \omega}^{ac}_{\mu} (-\partial_\nu D^{ab}_{\nu} ) \varphi^{bc}_{\mu}
\right) \;,  \label{exbrst}
\end{equation}
so that
\begin{equation}
S_\text{GZ} = S_\text{YM} + s \int d^4x \left( {\bar c}^a \partial_\mu
A^a_\mu + {\bar \omega}^{ac}_{\mu} (-\partial_\nu D^{ab}_{\nu} )
\varphi^{bc}_{\mu} \right) + S_{\gamma}\;,  \label{sgz1}
\end{equation}
from which eq.\eqref{brstbr} becomes apparent.

\section{The Polyakov loop and the background field formalism}
\label{Ploop}
In this section we shall investigate the confinement/deconfinement phase transition
of the $SU(2)$ gauge field theory in the presence of two static sources
of (heavy) quarks. The standard way
to achieve this goal is by probing the Polyakov loop order parameter,
\begin{eqnarray}
\mathcal{P} = \frac{1}{N}\tr \Braket{P e^{ig\int_{0}^{\beta}dt \;\;
A_{0}(t,x)}}\;,
\end{eqnarray}
with $P$ denoting path ordering, needed in the non-Abelian case to ensure
the gauge invariance of $\mathcal{P}$. This path ordering is not relevant at
one-loop order, which will considerably simplify the computations of the current work. In analytical studies of the phase transition involving the
Polyakov loop, one usually imposes the so-called ``Polyakov gauge'' on the
gauge field, in which case the time-component $A_{0}$ becomes diagonal and
independent of (imaginary) time. This means that the gauge field belongs to the Cartan subalgebra. More details on Polyakov gauge can be found in \cite{Marhauser:2008fz,Fukushima:2003fw,Ratti:2005jh}. Besides the trivial
simplification of the Polyakov loop, when imposing the Polyakov gauge it
turns out that the quantity $\Braket{A_{0}}$ becomes a good alternative choice for the order parameter instead of $\mathcal{P}$. This extra benefit can be proven by means of Jensen's
inequality for convex functions and is carefully explained in \cite{Marhauser:2008fz}, see also \cite{Braun:2007bx,Reinhardt:2012qe,Reinhardt:2013iia,Heffner:2015zna,Reinosa:2014ooa}. For
example, for the $SU(2)$ case we have the following: if $\frac{1}{2}g\beta%
\Braket{A_{0}} = \frac{\pi}{2}$ then we are in the ``unbroken symmetry phase''
(confined or disordered phase), equivalent to $\Braket{{\cal P}} = 0$;
otherwise, if $\frac{1}{2}g\beta\Braket{A_{0}} < \frac{\pi}{2}$, we are in the
``broken symmetry phase'' (deconfined or ordered phase), equivalent to
$\Braket{{\cal P}} \neq 0$. Since $\mathcal{P}\propto e^{-F T}$ with $T$ the temperature and $F$ the free energy of a heavy quark, it is clear that in the confinement phase, an infinite amount of energy would be required to actually get a free quark. The broken/restored symmetry referred to is the $\mathbb{Z}_N$ center symmetry of a pure gauge theory (no dynamical matter in the fundamental representation).

A slightly alternative approach to access the Polyakov loop was worked out in \cite{Reinosa:2014ooa}.
In order to probe the phase transition in a quantized non-Abelian gauge
field theory, we use, following \cite{Reinosa:2014ooa}, the Background Field Gauge (BFG) formalism, detailed in general in e.g.~\cite{Weinberg:1996kr}. Within this framework, the effective gauge field will be defined as the sum of a classical field $\bar{A}_{\mu}$ and a
quantum field $A_{\mu}$: $a_{\mu}(x) = a_{\mu}^{a}(x)t^{a} = \bar{A}_{\mu}+A_{\mu}
$, with $t^{a}$  the infinitesimal generators of the
$SU(N)$ symmetry group. The BFG method is a convenient approach, since the
tracking of breaking/restoration of the $\mathbb{Z}_{N}$ symmetry becomes
easier by choosing the Polyakov gauge for the background field.

Within this framework, it is convenient to define the gauge condition for the quantum field,
\begin{eqnarray}
\bar{D}_{\mu}A_{\mu} = 0\;,  \label{LDW}
\end{eqnarray}
 known as the Landau--DeWitt (LDW) gauge fixing condition, where $\bar{D}%
^{ab}_{\mu} =\delta^{ab}\partial_{\mu} - gf^{abc}\bar{A}^{c}_{\mu}$ is the
background covariant derivative. After integrating out the (gauge fixing)
auxiliary field $b^{a}$, we end up with the following Yang--Mills action,
\begin{eqnarray}
S_\text{BFG} = \int d^{d}x\; \left\{ \frac{1}{4}F^{a}_{\mu\nu}F^{a}_{\mu\nu}
- \frac{\left( \bar{D}A \right)^{2}}{2\xi} + \bar{c}^{a}\bar{D}_{\mu}^{ab}
D^{bd}_{\mu}(a)c^{d} \right\} \;.  \label{bfg}
\end{eqnarray}
Notice that, concerning the quantum field $A_{\mu}$, the condition %
\eqref{LDW} is equivalent to the Landau gauge, yet the action still has  background center symmetry. The LDW gauge is actually recovered in the limit
$\xi \to 0$, taken at the very end of each computation.

It is perhaps important here to stress that we are restricting our analysis to the (background) Landau gauge, for which a derivation argument in favor of the action \eqref{bfg} can be provided. For a vanishing background, this is precisely the original Gribov-Zwanziger construction \cite{Gribov:1977wm,Zwanziger:1989mf,Zwanziger:1992qr}, also applicable to the Coulomb gauge. More recently, it was also generalized to the $SU(2)$ maximal Abelian gauge in \cite{Gongyo:2013rua}. Intuitively, it might be clear that the precise influence on the quantum dynamics by Gribov copies can strongly depend on the chosen background, given that Gribov copies are defined via the zero modes of the Faddeev-Popov operator of the chosen gauge condition, which itself explicitly depends on the chosen background. This is open to further research, as it has not been pursued in the literature yet. Though, for a constant background as relevant for the current purposes, it will be discussed elsewhere that the action is indeed obtainable via a suitable extension of the arguments of \cite{Gribov:1977wm,Zwanziger:1989mf,Zwanziger:1992qr}.

In the absence of a background, a proposal for a generalization to the linear covariant gauges was put forward in \cite{Lavrov:2011wb,Lavrov:2013boa}, albeit leading to a very complicated nonlocal Lagrangian structure, containing e.g.~reciprocals and exponentials of fields. To our knowledge, no practical computations were done so far with this formalism. Nonetheless, potential problems with gauge parameter dependence of physical quantities were discussed in \cite{Lavrov:2011wb,Lavrov:2013boa}, not surprisingly linked to the softly broken BRST symmetry, see also our Section 2 for more on this and relevant references.

A very recent alternative for the linear covariant gauges was worked out in \cite{Capri:2015pja}, partially building on earlier work of \cite{Sobreiro:2005vn}.  With this proposal, it was explicitly checked at one loop that the Gribov parameter $\gamma^2$ and vacuum energy are gauge parameter independent. This at least suggests that in this class of covariant gauges, an approach to Gribov copies can be worked out that is compatible with gauge parameter independence \cite{Capri:2015pjab}.

As explained for the simple Landau gauge in the previous section, the Landau
background gauge condition is also plagued by Gribov ambiguities, and the
Gribov--Zwanziger procedure is applicable also in this instance. The starting
point of our analysis is, therefore, the GZ action modified for the BFG
framework (see \cite{Zwanziger:1982na}):
\begin{multline}
S_\text{GZ+PLoop} = \int d^{d}x\; \left\{ \frac{1}{4}F^{a}_{\mu\nu}F^{a}_{%
\mu\nu} - \frac{\left( \bar{D}A \right)^{2}}{2\xi} + \bar{c}^{a}\bar{D}%
_{\mu}^{ab}D^{bd}_{\mu}(a)c^{d} + \bar{\varphi}_{\mu}^{ac} \bar{D}%
_{\nu}^{ab}D^{bd}_{\nu}(a) \varphi_\mu^{dc} \right. \\
\left. - \bar{\omega}_{\mu}^{ac} \bar{D}_{\nu}^{ab}D^{bd}_{\nu}(a)
\omega_\mu^{dc} - g\gamma ^{2} f^{abc}A_\mu^a \left( \varphi_\mu^{bc} + \bar{%
\varphi}_\mu^{bc} \right) - \gamma^{4}d(N^{2}-1) \right\}\;.  \label{gzpl}
\end{multline}
As mentioned before, with the Polyakov gauge imposed to the background
field $\bar{A}_{\mu}$, the time-component becomes diagonal and
time-independent. In other words, we have $\bar{A}_{\mu}(x) = \bar{A}%
_{0}\delta_{\mu 0}$, with $\bar{A}_{0}$ belonging to the Cartan subalgebra
of the gauge group. For instance, in the Cartan subalgebra of $SU(2)$ only
the $t^{3}$ generator is present, so that $\bar{A}^{a}_{0} = \delta^{a3}\bar{%
A}^{3}_{0}\equiv \delta^{a3}\bar A_0$. As explained in \cite{Reinosa:2014ooa}, at leading order we then simply find, using the properties of the Pauli matrices,
\begin{equation}
\mathcal{P}=\cos\frac{r}{2}\,,
\end{equation}
where we defined
\begin{equation}
  r=g\beta \bar{A}_0\,,
\end{equation}
with $\beta$ the inverse temperature. Just like before, $r=\pi$ corresponds to the confinement phase, while $0\leq r<\pi$ corresponds to deconfinement. With a slight abuse of language, we will refer to the quantity $r$ as the Polyakov loop hereafter.

Since the scope of this work is limited to one-loop order, only terms quadratic in the quantum fields in the action \eqref{gzpl} shall be considered. One then immediately gets an action that can be split in term coming from the two color sectors: the 3rd color direction, called Cartan direction, which does not depend on the parameter $r$; and one coming from the $2\times 2$ block given by the 1st and 2nd color directions. This second $2\times2$ color sector is orthogonal to the Cartan direction and does depend on $r$. The scenario can then be seen as a system where the vector field has an imaginary chemical potential $i rT$ and has isospins $+1$ and $-1$ related to the $2\times2$ color sector and one isospin $0$ related to the $1\times1$ color sector.

\section{The finite temperature effective action at leading order}
\label{sec4}

Considering only the quadratic terms of \eqref{gzpl}, the integration of the
partition function gives us the following vacuum energy at one-loop order, defined according
to \eqref{vce},
\begin{equation}
\beta V\mathcal{E}_{v}=-\frac{d(N^{2}-1)}{2Ng^{2}}\lambda ^{4}+\frac{1}{2}%
(d-1)\tr\ln \frac{D^{4}+\lambda ^{4}}{-D^{2}}-\frac{1}{2}%
\tr\ln (-D^{2})\;,  \label{vace}
\end{equation}%
where $V$ is now just the spacial volume. Here, $\mathcal{D}$ is the covariant derivative in the adjoint
representation in the presence of the background $A_{0}^{3}$ field and $%
\lambda ^{4}=2Ng^{2}\gamma ^{4}$. Throughout this work, it is always tacitly
assumed we are working with $N=2$ colors, although we will frequently
continue to explicitly write $N$ dependence for generality. Using the usual Matsubara formalism, we have that $%
\mathcal{D}^{2}=(2\pi nT+rsT)^{2}+\vec{q}^{2}$, where $n$ is the Matsubara
mode, $\vec{q}$ is the spacelike momentum component, and $s$ is the isospin,
given by $-1$, $0$, or $+1$ for the $SU(2)$ case\footnote{%
The $SU(3)$ case was handled in \cite{Reinosa:2014ooa} as well (see also \cite{Serreau:2015saa}).}.

The general trace is of the form
\begin{equation}  \label{1}
\frac{1}{\beta V}\tr\ln (-D^2 + m^2) = T \sum_s
\sum_{n=-\infty}^{+\infty}\int\frac{ d^{3-\epsilon}q}{(2\pi)^{3-\epsilon}}
\ln \left((2\pi nT + rsT)^2+\vec{q}^2+m^2\right)\,,
\end{equation}
which will be computed immediately below.

\subsection{The sum-integral: 2 different computations}
We want to compute the following expression:
\begin{equation}
\mathcal{I }= T \sum_{n=-\infty}^{+\infty}\int\frac{ d^{3-\epsilon}q}{%
(2\pi)^{3-\epsilon}} \ln \left((2\pi nT + rT)^2+\vec{q}^2+m^2\right) \;.
\end{equation}
One way to proceed is to start by deriving the previous expression with
respect to $m^2$. Then, one can use the well-known formula from complex
analysis
\begin{equation}  \label{3}
\sum_{n=-\infty}^{+\infty} f(n) = -\pi\sum_{z_0}\mathop{\mathcal Res}%
\limits_{z=z_0}\cot(\pi z)f(z)
\end{equation}
where the sum is over the poles $z_0$ of the function $f(z)$. Subsequently
we integrate with respect to $m^2$ (and determine the integration
constant by matching the result with the known $T=0$ case). Finally one can split off the analogous $T=0$ trace (which does not depend on the background
field) to find
\begin{equation}  \label{6}
\mathcal{I }= \int \frac{d^{4-\epsilon}q}{(2\pi)^{4-\epsilon}}\ln(q^2+m^2) +
T \int\frac{d^{3}q}{(2\pi)^{3}}\ln\left(1+e^{-2\frac{\sqrt{\vec{q}^2+m^2}}{T}%
}-2e^{-\frac{\sqrt{\vec{q}^2+m^2}}{T}}\cos r\right) \;.
\end{equation}
where the limit $\epsilon\to0$ was taken in the (convergent) second
integral. The first term in the r.h.s.~is the (divergent) zero temperature
contribution.

Another way to compute the above integral is by making use of Zeta function regularization techniques, which are particularly useful in the computation  of the Casimir energy in various configurations see \cite{elizalde95,bordag10}. The
advantage of this second technique is that, although it is less direct, it
provides one with an easy way to analyze the high and low temperature limits as well as the small mass limit, as we will now show. Moreover, within
this framework, the regularization procedures are often quite
transparent. One starts by writing the logarithm as $\ln
x=-\lim_{s\rightarrow 0}\partial _{s}x^{-s}$, after which the integral over
the momenta can be performed:
\begin{equation}
\mathcal{I}=-T\lim_{s\rightarrow 0}\partial _{s}\left( \mu
^{2s}\sum_{n=-\infty }^{\infty }\frac{\Gamma (s-3/2)}{8\pi ^{\frac{3}{2}%
}\Gamma (s)}\left[ (2\pi nT+rT)^{2}+m^{2}\right] ^{\frac{3}{2}-s}\right) \;,
\end{equation}%
where the renormalization scale $\mu $ has been introduced to get
dimensional agreement for $s\not=0$, and where we already put $\epsilon =0$, as $s$ will function as a regulator --- i.e.~we assume $s>3/2$ and
analytically continuate to bring $s\rightarrow 0$. Using the integral
representation of the Gamma function, the previous expression can be recast
to
\begin{eqnarray}
\mathcal{I} &=&-T\lim_{s\rightarrow 0}\partial _{s}\left( \mu
^{2s}\sum_{n=-\infty }^{\infty }\frac{1}{8\pi ^{\frac{3}{2}}\Gamma (s)}%
\int_{0}^{\infty }t^{s-5/2}e^{-t\left( (2\pi nT+rT)^{2}+m^{2}\right)
}dt\right)   \notag \\
&=&-\lim_{s\rightarrow 0}\partial _{s}\left( \mu ^{2s}\frac{T^{4-2s}}{%
4^{s}\pi ^{2s-3/2}\Gamma (s)}\int_{0}^{\infty }dyy^{s-5/2}e^{-\frac{m^{2}y%
}{4\pi ^{2}T^{2}}}\sum_{n=-\infty }^{\infty }e^{-y(n+\frac{r}{2\pi }%
)^{2}}\right) \;,
\end{eqnarray}%
where the variable of integration was transformed as $y=4\pi ^{2}T^{2}t\geq0 $ in
the second line. Using the Poisson rule (valid for positive $\omega $):
\begin{equation}
\sum_{n=-\infty }^{+\infty }e^{-(n+x)^{2}\omega }=\sqrt{\frac{\pi }{\omega }}%
\left( 1+2\sum_{n=1}^{\infty }e^{-\frac{n^{2}\pi ^{2}}{\omega }}\cos {(2n\pi
x)}\right) \;,
\end{equation}%
we obtain that
\begin{multline}
\mathcal{I}=-\lim_{s\rightarrow 0}\partial _{s}\mu ^{2s}\left[ \frac{\Gamma
(s-2)T^{4-2s}}{4^{s}\pi ^{2s-2}\Gamma (s)}\left( \frac{m^{2}}{4\pi ^{2}T^{2}}%
\right) ^{2-s}+\right.  \\
\left. \frac{T^{4-2s}}{4^{s-1}\pi ^{s}\Gamma (s)}\left( \frac{m^{2}}{4\pi
^{2}T^{2}}\right) ^{1-s/2}\sum_{n=1}^{\infty }{n^{s-2}\cos {(nr)}%
K_{2-s}\left( \frac{nm}{T}\right) }\right] \;,
\end{multline}%
where $K_{\nu }(z)$ is the modified Bessel function of the second kind.
Simplifying this, we find
\begin{equation}
\mathcal{I}=\frac{m^{4}}{2(4\pi )^{2}}\left[ \ln {\left( \frac{m^{2}}{\mu
^{2}}\right) }-\frac{3}{2}\right] -\sum_{n=1}^{\infty }{\frac{m^{2}T^{2}\cos
{(nr)}}{\pi ^{2}n^{2}}K_{2}\left( \frac{nm}{T}\right) }\;,  \label{result2}
\end{equation}%
where the first term is the $T=0$ contribution, and the sum is the
finite-temperature correction. Using numerical integration and series
summation, it can be checked that both results \eqref{6} and \eqref{result2}
are indeed identical. Throughout this paper, we will mostly base ourselves
on the expression \eqref{6}. Nonetheless the Bessel series is quite useful in
obtaining the limit cases $m=0$, $T\rightarrow \infty $, and $T\rightarrow 0$
by means of the corresponding behaviour of $K_{2}(z)$. Observing that
\begin{equation}
\lim_{m\rightarrow 0}\left( -\frac{m^{2}T^{2}K_{2}\left( \frac{mn}{T}\right)
\cos (nrs)}{\pi ^{2}n^{2}}\right) =-\frac{2T^{4}\cos (nrs)}{\pi ^{2}n^{4}}\,,
\end{equation}%
we obtain
\begin{equation}
\mathcal{I}_{m=0}=-\frac{T^{4}}{\pi ^{2}}\left[ \text{Li}_{4}\left(
e^{-irs}\right) +\text{Li}_{4}\left( e^{irs}\right) \right],
\end{equation}%
where $\text{Li}_{s}(z)=\sum_{n=1}^{\infty}\frac{z^{n}}{n^{s}}$ is the polylogarithm or Jonqui\`ere's function.

Analogously,
\begin{equation*}
\lim_{T\rightarrow\infty}K_2\left( \frac{m n}{T}\right)\sim \frac{2 T^2}{m^2 n^2}-\frac{1}{2}\,,
\end{equation*}
so that
\begin{equation}
\mathcal{I}_{T\rightarrow\infty} = \frac{m^4}{2(4\pi)^2}\left[ \ln{\left(
\frac{m^2}{\mu^2}\right)}-\frac{3}{2} \right]+ \frac{T^2}{4 \pi^2 }
\left\{m^2 \left[\text{Li}_2\left(e^{-i r s}\right)+\text{Li}_2\left(e^{i r
s}\right)\right]-4 T^2 \left[\text{Li}_4\left(e^{-i r s}\right)+\text{Li}%
_4\left(e^{i r s}\right)\right]\right\}\,.
\end{equation}
Finally for $T \rightarrow 0$ we can use the asymptotic expansion of the
Bessel function \cite{as}:
\begin{equation}
K_\nu(z)\sim\sqrt{\frac{\pi}{2z}}e^{-z}\left( \sum_{k=0}^{\infty}\frac{%
a_k(\nu)}{z^k}\right),\ \ |\text{Arg}( z)|\leq\frac{3}{2}\pi\,,
\end{equation}
where $a_{k}(\nu)$ are finite factors. So, at first order ($k=0$),
\begin{eqnarray}
\mathcal{I}_{T\rightarrow0} = \frac{m^4}{2(4\pi)^2}\left[ \ln{\left( \frac{%
m^2}{\mu^2}\right)}-\frac{3}{2} \right]-\frac{m^{3/2} T^{5/2} }{2 \sqrt{2}
\pi ^{3/2}} \left[\text{Li}_{\frac{5}{2}}\left(e^{-\frac{m}{T}-i r s}\right)+%
\text{Li}_{\frac{5}{2}}\left(e^{-\frac{m}{T}+i r s}\right)\right]\,.
\end{eqnarray}

\subsection{The result for further usage}
Making use of the result \eqref{6} we may define
\begin{equation}  \label{defi}
I(m^2,r,s,T) = T \int\frac{d^{3}q}{(2\pi)^{3}}\ln\left(1+e^{-2\frac{\sqrt{%
\vec{q}^2+m^2}}{T}}-2e^{-\frac{\sqrt{\vec{q}^2+m^2}}{T}}\cos rs\right),
\end{equation}
so that the vacuum energy \eqref{vace} can be rewritten as
\begin{equation}  \label{vactwee}
\begin{aligned} {\cal E}_{v} = &- \frac{d(N^2-1)}{2Ng^2} \lambda^4 + \frac12
(d-1)(N^2-1) \tr_{T=0}\ln\frac{\partial^4+\lambda^4}{-\partial^2} - \frac12
(N^2-1) \tr_{T=0}\ln(-\partial^2) \\ &+ \sum_s \left( \frac12 (d-1)
(I(i\lambda^2,r,s,T)+I(-i\lambda^2,r,s,T)-I(0,r,s,T)) - \frac12 I(0,r,s,T)
\right) \;, \end{aligned}
\end{equation}
where $\tr_{T=0}$ denotes the trace taken at zero temperature.

\section{Minimization of the effective action, the Polyakov loop and the
Gribov mass}

\subsection{Warming-up exercise: assuming a $T$-independent Gribov mass $\lambda$}
\label{Tonafhankelijk}

\begin{figure}[tbp]
\begin{center}
\includegraphics[width=.5\textwidth]{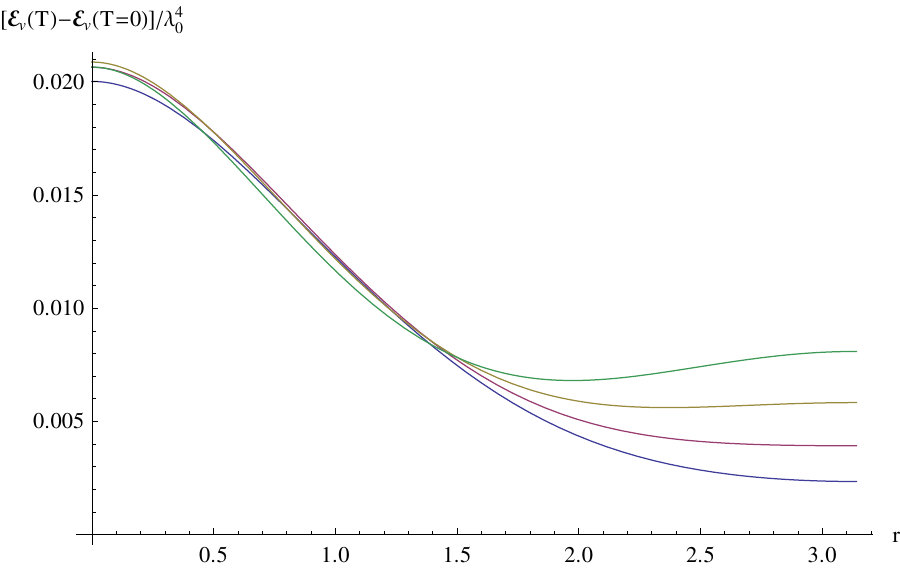}
\end{center}
\caption{The effective potential \eqref{vactwee} at the temperatures (from
below upwards at $r=\protect\pi$) 0.42, 0.44, 0.46, and 0.48 times $\protect\lambda$ as a function of $r$, with the simplifying assumption that $\protect%
\lambda$ maintains its zero-temperature value $\lambda_0$ throughout. It can be seen that
the minimum of the potential moves away from $r=\protect\pi$ in between $T=%
\unit{0.44}{\protect\lambda}$ and $\unit{0.46}{\protect\lambda}$. }
\label{energieplot}
\end{figure}

As a first simpler case, let us simplify matters slightly by assuming that the temperature does not influence the Gribov parameter $\lambda$. This means that $\lambda$ will be supposed to assume its zero-temperature value, which we will call $\lambda_{0}$, given by the solution of the gap equation \eqref{hc1} at zero temperature. In this case, only the terms with the function $I$ matter in \eqref{vactwee}, since the other terms do not explicitely depend on the Polyakov line $r$. Plotting this part of the potential (see \figurename\ \ref{energieplot}), one finds by visual inspection that a second-order phase transition occurs from the minimum with $r=\pi$ to a minimum with $r\not=\pi$. The transition can be identified by the condition\begin{equation}  \label{dvdr}
\left. \frac{d^2}{dr^2} \mathcal{E}_{v} \right|_{r=\pi} = 0 \;.
\end{equation}
Using the fact that
\begin{equation}
\frac{\partial^2I}{\partial r^2}(m^2,r=\pi,s,T) = -2T \int\frac{d^{3}q}{%
(2\pi)^{3}}\frac{e^{-\frac{\sqrt{\vec{q}^2+m^2}}{T}}}{\left(1+e^{-\frac{%
\sqrt{\vec{q}^2+m^2}}T}\right)^2}
\end{equation}
when $s=\pm1$ and zero when $s=0$, the equation \eqref{dvdr} can be
straightforwardly solved numerically for the critical temperature. We find
\begin{equation}  \label{Tcrit}
T_\text{crit} = \unit{0.45}{\lambda}_{0} \;.
\end{equation}

\subsection{The $T$-dependence of the Gribov mass $\lambda$}

\label{gammavanT} Let us now investigate what happens to the Gribov
parameter $\lambda$ when the temperature is nonzero. Taking the derivative of
the effective potential \eqref{vactwee} with respect to $\lambda^2$ and
dividing by $d(N^2-1)\lambda^2/Ng^2$ (as we are not interested in the solution $\lambda^2=0$) yields the gap equation for general number of colors $N$:
\begin{equation}  \label{gapeen}
1 = \frac12 \frac{d-1}d Ng^2 \tr \frac1{\partial^4+\lambda^4} + \frac12
\frac{d-1}d \frac{Ng^2}{N^2-1} \frac i{\lambda^2} \sum_s \left(\frac{%
\partial I}{\partial m^2}(i\lambda^2,r,s,T) - \frac{\partial I}{\partial m^2}%
(-i\lambda^2,r,s,T)\right) \;,
\end{equation}
where the notation $\partial I/\partial m^2$ denotes the derivative of $I$
with respect to its first argument (written $m^2$ in \eqref{defi}). If we
now define $\lambda_0$ to be the solution to the gap equation at $T=0$:
\begin{equation}
1 = \frac12 \frac{d-1}d Ng^2 \tr \frac1{\partial^4+\lambda_0^4} \;,
\label{t0gapeq}
\end{equation}
then we can subtract this equation from the general gap equation %
\eqref{gapeen}. After dividing through $(d-1)Ng^2/2d$ and setting $d=4$ and $%
N=2$, the result is
\begin{equation}
\int \frac{d^4q}{(2\pi)^4} \left(\frac1{q^4+\lambda^4}-\frac1{q^4+%
\lambda_0^4}\right) + \frac i{3\lambda^2} \sum_s \left(\frac{\partial I}{%
\partial m^2}(i\lambda^2,r,s,T) - \frac{\partial I}{\partial m^2}%
(-i\lambda^2,r,s,T)\right) = 0 \;,
\end{equation}
where now all integrations are convergent. This equation can be easily
solved numerically to yield $\lambda$ as a function of temperature $T$ and
background $r$, in units $\lambda_0$. This is shown in \figurename\ \ref%
{gammaplot}.

\begin{figure}[tbp]
\begin{center}
\includegraphics[width=.5\textwidth]{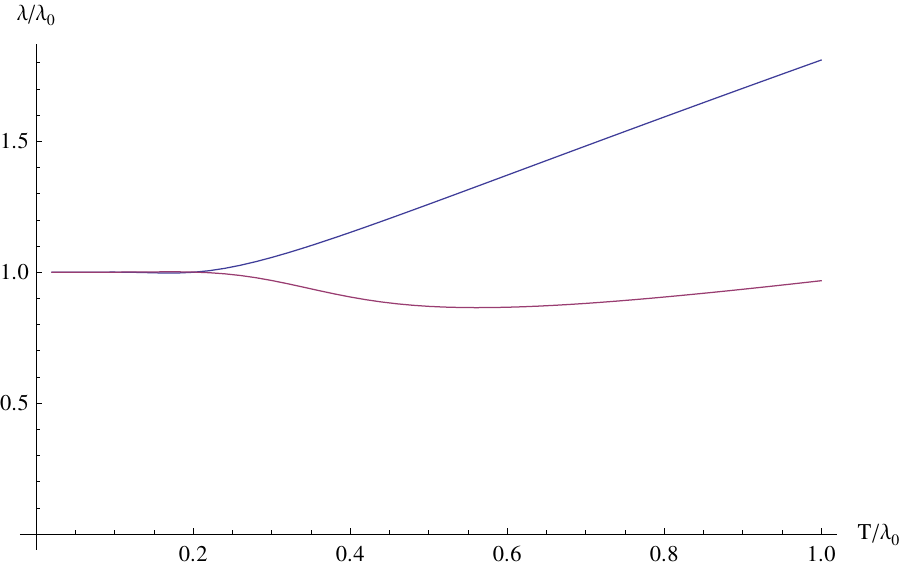}
\end{center}
\caption{The Gribov parameter $\protect\lambda$ as a function of the
temperature $T$ at $r$ equals to zero (upper line) and $\protect\pi$ (lower
line), in units of the zero-temperature Gribov parameter $\protect\lambda_0$.
}
\label{gammaplot}
\end{figure}

\subsection{Absolute minimum of the effective action}

As $\lambda$ does not change much when including its dependence on
temperature and background, the transition is still second order and its
temperature is, therefore, still given by the condition \eqref{dvdr}. Now,
however, the potential depends explicitely on $r$, but also implicitely due
to the presence of the $r$-dependent $\lambda$. We therefore have
\begin{equation}
\left. \frac{d^2}{dr^2} \mathcal{E}_{v} \right|_{r=\pi} = \left. \frac{%
\partial^2\mathcal{E}_{v}}{\partial r^2} + 2 \frac{d\lambda}{dr} \frac{%
\partial^2\mathcal{E}_{v}}{\partial r\partial\lambda} + \frac{d^2\lambda}{%
dr^2} \frac{\partial \mathcal{E}_{v}}{\partial\lambda} + \left(\frac{d\lambda%
}{dr}\right)^2 \frac{\partial^2\mathcal{E}_{v}}{\partial\lambda^2}%
\right|_{\lambda=\lambda(r),r=\pi} \;.
\end{equation}
Now, $d\lambda/dr|_{r=\pi} = 0$ due to the symmetry at that point.
Furthermore, as we are considering $\lambda\neq0$, $\partial \mathcal{E}_{v}/\partial\lambda = 0$ is the gap
equation and is solved by $\lambda(r)$. Therefore, we find for the condition
of the transition:
\begin{equation}  \label{dvdrpartial}
\left. \frac{\partial^2\mathcal{E}_{v}}{\partial r^2} (r,\lambda,T) \right|_{r=\pi} = 0 \;,
\end{equation}
where the derivative is taken with respect to the explicit $r$ only.

We already solved equation \eqref{dvdrpartial} in section \ref%
{Tonafhankelijk}, giving \eqref{Tcrit}:
\begin{equation}
T = 0.45 \lambda(r,T) \;.
\end{equation}
As we computed $\lambda$ as a function of $r$ and $T$ in section \ref%
{gammavanT} already, it is again straightforward to solve this equation to
give the eventual critical temperature:
\begin{equation}
T_\text{crit} = \unit{0.40}{\lambda_0} \;,
\end{equation}
as expected only slightly different from the simplified estimate \eqref{Tcrit} found before. 

\subsection{The $T$-dependence of the Polyakov loop $r$ and the equation of
state}

\subsubsection{Deconfinement transition and its imprint on the Gribov mass}

\begin{figure}[h]
\begin{center}
\includegraphics[width=.5\textwidth]{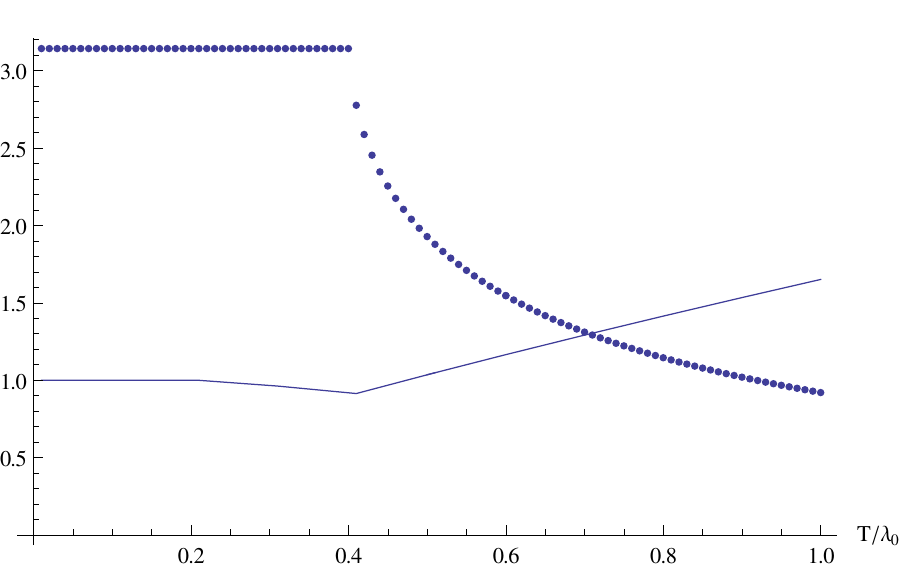}
\end{center}
\caption{The dotted line curve represents $r(T)$, while the continuous line is $\protect\lambda(T)$. At $T\approx 0.40\protect\lambda_{0}$, both curves clearly have a discontinuous derivative.}
\label{randlambda}
\end{figure}

Let us now investigate the temperature dependence of $r$. The physical value of the background field $r$ is found by minimizing the vacuum energy:
\begin{eqnarray}
\frac{d}{dr} {\cal E}_{v} =0\;.
\end{eqnarray}
From the vacuum energy  \eqref{vactwee} we have
\begin{eqnarray}
\frac{\partial\mathcal{E}_{v}}{\partial r} = (d-1) \left[ \frac{\partial
I}{\partial r}(i\gamma^2,r,T) + \frac{\partial I}{\partial r}(-i\gamma^2,r,T)
- \frac{d}{(d-1)} \frac{\partial I}{\partial r}(0,r,T) \right] = 0\;.
\label{rgapeq}
\end{eqnarray}
The expression \eqref{rgapeq} was obtained after  summation over the
possible values of $s$. Furthermore, we used the fact that $%
I(m^{2},r,+1,T)=I(m^{2},r,-1,T)$ and that $s=0$ accounts for terms
independent of $r$, which are cancelled by the derivation w.r.t. $r$. From %
\eqref{defi} one can get, whenever $s=\pm1$:
\begin{eqnarray}
\frac{\partial I(m^{2},r,T)}{\partial r} = T \int\frac{d^{3}q}{(2\pi)^{3}}
\frac{2e^{-\frac{\sqrt{\vec{q}^2+m^2}}{T}}\sin r}{\left(1+e^{-2\frac{\sqrt{%
\vec{q}^2+m^2}}{T}}-2e^{-\frac{\sqrt{\vec{q}^2+m^2}}{T}}\cos r \right)}\;.
\end{eqnarray}
Since \eqref{rgapeq} is finite, we can numerically obtain $r$ as a function of
temperature. From the dotted curve in Figure \ref{randlambda} one can easily see that, for $T > T_{\text{crit}} \approx 0.40 \lambda_{0}$, we have $%
r \neq \pi$, pointing to a deconfined phase,  confirming
the computations of the previous section. In the same figure, $\lambda(T)$ is plotted in a continuous line. We observe very clearly
that the Gribov mass $\lambda(T)$ develops a cusp-like behaviour exactly at
the critical temperature $T =T_{\text{crit}}$.

\subsubsection{Equation of state}

\begin{figure}[h]
\begin{center}
\includegraphics[width=.45\textwidth]{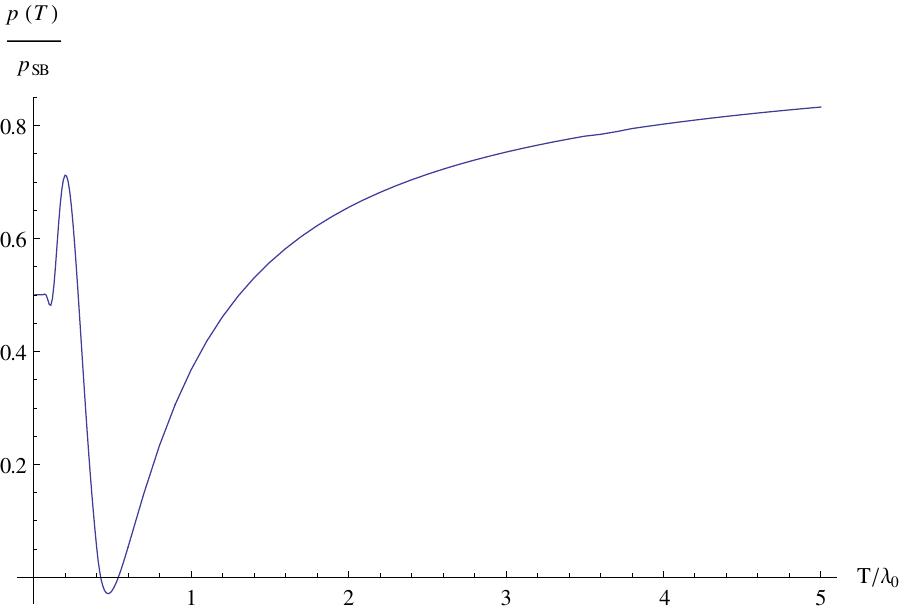} \hspace{10mm} %
\includegraphics[width=.45\textwidth]{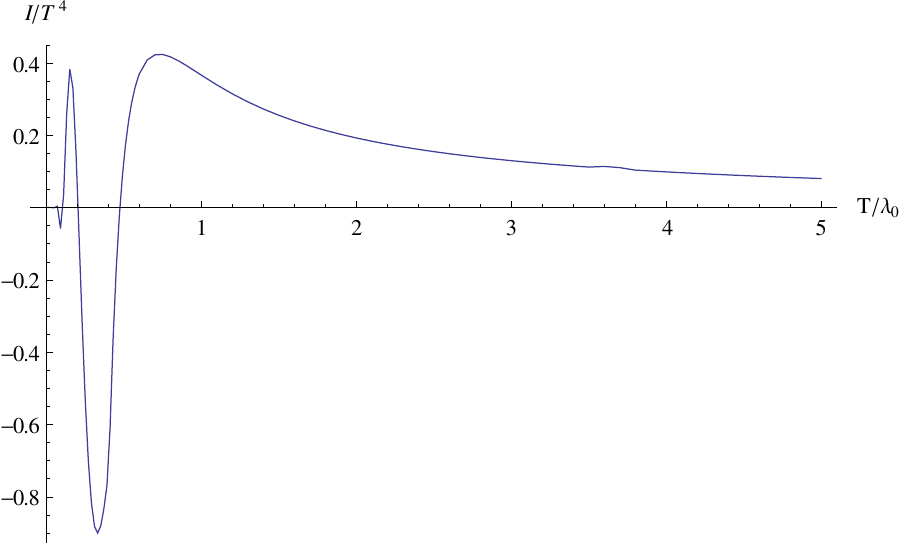}
\end{center}
\caption{Left: GZ pressure (relative to the Stefan-Boltzmann limit pressure $%
\sim T^{4}$). Right: GZ trace anomaly.}
\label{PTraceAnom}
\end{figure}

Following \cite{Philipsen:2012nu}, we can also extract an estimate for the
(density) pressure $p$ and the interaction measure $I/T^{4}$, shown in
Figure \ref{PTraceAnom} (left and right  respectively). As usual the
(density) pressure is defined as
\begin{eqnarray}
p = \frac{1}{\beta V}\ln Z_{GZ}\;,
\end{eqnarray}
which is related to the free energy by $p = -\mathcal{E}_{v}$. Here the plot of the pressure is given relative to the Stefan--Boltzmann limit pressure: $p_{SB} = \kappa T^{4}$, where $\kappa = (N^2 -1)\pi T^{4}/45$ is the Stefan--Boltzmann constant accounting for all degrees of freedom of the system at high temperature. We subtract the zero-temperature value, such that the pressure becomes zero at zero temperature: $p(T) = - [\mathcal{E}_{v}(T) - \mathcal{E}_{v}(T=0)]$. Namely, after using the $\MSbar$ renormalization prescription and choosing the renormalization parameter $\bar{\mu}$ so that the zero temperature gap equation is satisfied,
\begin{eqnarray}
{\bar{\mu}}^2 = \lambda^{2}_{0}e^{-\left(  \frac{5}{6} - \frac{32\pi^2}{3g^{2}} \right)}\;,
\end{eqnarray}
we have the following expression for the pressure (in units of $\lambda_{0}^{4}$),
\begin{eqnarray}
-\frac{p(T)}{\lambda_{0}^{4}} &=& 3 \left[  I(i\lambda'^{2},r,T') + I(-i\lambda'^{2},r,T') - \frac{4}{3}I(0,r,T')\right]
\nonumber \\
&+&
\frac{3}{2} \left[ I(i\lambda'^{2},0,T') + I(-i\lambda'^{2},0,T') - \frac{4}{3}I(0,0,T')\right]
\nonumber \\
&-&
\frac{9\lambda'^{4}}{32\pi^2 } \left(  \ln \lambda'^{2}  - \frac12 \right) - \frac{9}{64\pi^{2}}
\;. \label{vcen1}
\end{eqnarray}
In \eqref{vcen1} prime quantities stand for quantities in units of $\lambda_{0}$, while $\lambda$ and $\lambda_{0}$ satisfy their gap equation. The last term of \eqref{vcen1} accounts for the zero temperature subtraction, so that $p(0) = 0$, according to the definition of $I(m^2, r,T)$ in \eqref{defi}. Note that the coupling constant does not explicitly appear in \eqref{vcen1} and that $\lambda_{0}$ stands for the Gribov parameter at $T=0$.

The interaction measure $I$ is defined as the trace anomaly in units of $T^{4}$, and $I$ is nothing less than the trace of the of the stress-energy
tensor, given by
\begin{eqnarray}
\theta_{\mu\nu} = (p + \epsilon)u_{\mu}u_{\nu} - p\eta_{\mu\nu}\;,
\end{eqnarray}
with $\epsilon$ being the internal energy density, which is defined as $\epsilon = {\cal E}_{v} + Ts$ (with $s$ the entropy density), $u = (1,0,0,0)$ and $\eta_{\mu\nu}$ the (Euclidean) metric of the space-time. Given the thermodynamic definitions of each quantity (energy, pressure and entropy), we obtain
\begin{eqnarray}
I = \theta_{\mu\mu} = T^{5}\frac{\partial}{\partial T}\left(\frac{p}{T^{4}}\right)\;.
\end{eqnarray}
Both quantities display a behavior similar to that presented in \cite{Fukushima:2013xsa} (but note that in they plot the temperature in units of the critical temperature ($T_{c}$ in their notation), while we use units $\lambda_{0}$). Besides this, and the fact that we included the effect of Polyakov loop on the Gribov parameter, in \cite{Fukushima:2013xsa} a lattice-inspired effective coupling was introduced at finite temperature while we used the exact one-loop perturbative expression, which is consistent with the order of all the computations made here.

However, we notice that at temperatures relatively close to our $T_{c}$, the
pressure becomes negative. This is clearly an unphysical feature, possibly
related to some missing essential physics. For higher temperatures, the
situation is fine and the pressure moreover displays a behaviour similar to what is seen
in lattice simulations for the nonperturbative pressure (see \cite{Borsanyi:2012ve} for the $SU(3)$ case). A similar problem is present in one
of the plots presented in \cite[Fig.~4]{Fukushima:2013xsa}, although no comment is made about it. Another strange feature is the oscillating behaviour of both
pressure and interaction measure at low temperatures. Something similar was already observed in \cite{Benic:2012ec} where a quark model was employed with complex conjugate quark mass. It is well-known that the gluon propagator develops two complex conjugate masses in Gribov--Zwanziger quantization, see e.g.~\cite{Dudal:2010cd,Dudal:2013wja,Baulieu:2009ha,Cucchieri:2011ig} for some more details, so we confirm the findings of \cite{Benic:2012ec} that, at least at leading order, the thermodynamic quantities develop an oscillatory behaviour. We expect this oscillatory behaviour would in principle also be present in \cite{Fukushima:2013xsa} if the pressure and interaction energy were to be computed at lower temperatures than shown there. In any case, the presence of complex masses and their consequences gives us a warning that a certain care is needed when using GZ dynamics, also at the level of spectral properties as done in \cite{Su:2014rma,Florkowski:2015rua}, see also \cite{Baulieu:2008fy,Dudal:2010wn}.

These peculiarities justify giving an outline in the next Section of the behaviours of the pressure and interaction measure in an improved formalism, such as in the Refined Gribov-Zwanziger one.

\section{Outlook to the Refined Gribov--Zwanziger formalism}

The previous results can be slightly generalized to the case of the Refined
Gribov--Zwanziger (RGZ) formalism studied in \cite%
{Dudal:2007cw,Dudal:2008sp,Dudal:2011gd,Gracey:2010cg,Thelan:2014mza}. In
this refined case, additional nonperturbative vacuum condensates such as $%
\langle A_\mu^2\rangle$ and $\langle\bar\varphi_\mu^{ab}\varphi_\mu^{ab}%
\rangle$ are to be introduced. The corresponding mass dimension two
operators get a nonzero vacuum expectation value (thereby further lowering
the vacuum energy) and thus influence the form of the propagator and
effective action computation. The predictions for the RGZ propagators, see
also \cite{Dudal:2010tf,Dudal:2012zx,Oliveira:2012eh}, are in fine agreement
with ruling $T=0$ lattice data, see e.g.~also \cite%
{lattice1,lattice2,lattice3,lattice4,lattice5,lattice6,lattice7,lattice8,lattice9}. This is in contrast with the original GZ predictions, such that it could happen that the finite temperature version of RGZ is also better suited to describe the phase transition and/or thermodynamical properties of
the pure gauge theory.

Due to the more complex nature of the RGZ effective action (more vacuum
condensates), we will relegate a detailed (variational) analysis of their
finite temperature counterparts\footnote{From \cite{Chernodub:2008kf,Dudal:2009tq,Vercauteren:2010rk}, the nontrivial response of the $d=2$ condensate $\braket{A^2}$ to temperature already became clear.} to future work, as this will require
new tools. Here, we only wish to present a first estimate of the
deconfinement critical temperature $T_c$ using as input the $T=0$ RGZ gluon
propagator where the nonperturbative mass parameters are fitted to lattice
data for the same propagator. More precisely, we use \cite{Cucchieri:2011ig}
\begin{equation}
\Delta_{\mu\nu}^{ab}(p) = \delta^{ab} \frac{p^2+M^2+\rho_1}{%
p^4+p^2(M^2+m^2+\rho_1) + m^2(M^2+\rho_1)+\lambda^4} \left(\delta_{\mu\nu} -
\frac{p_\mu p_\nu}{p^2}\right) \;.
\label{rgzgluonprop}
\end{equation}
where we omitted the global normalization factor $Z$ which drops out from
our leading order computation\footnote{This $Z$ is related to the choice of a MOM renormalization scheme, the kind
of scheme that can also be applied to lattice Green functions, in contrast
with the $\overline{\mbox{MS}}$ scheme.}. In this expression, we have that
\begin{equation}
\langle A_\mu^aA_\mu^a\rangle \to -m^2 \;, \qquad
\langle\bar\varphi_\mu^{ab}\varphi_\mu^{ab}\rangle \to M^2 \;, \qquad
\frac12 \langle \varphi_\mu^{ab}\varphi_\mu^{ab} +
\bar\varphi_\mu^{ab}\bar\varphi_\mu^{ab} \rangle \to \rho_1 \;.
\end{equation}
The free energy associated to the RGZ framework can be obtained by following the same steps as in section \ref{sec4}, leading to
\begin{eqnarray}
{\cal E}_{v}(T) &=& (d-1)\left[ I(r_{+}^{2}, r,T) + I(r_{-}^{2}, r,T) - I(N^{2}, r,T) - \frac{1}{d-1}I(0, r,T) \right]
\nonumber \\
&&
+\frac{(d-1)}{2}\left[ I(r_{+}^{2}, 0,T) + I(r_{-}^{2}, 0,T) - I(N^{2}, 0,T) - \frac{1}{d-1}I(0, 0,T) \right]
\nonumber \\
&&
+\int \frac{d^{d}p}{ (2\pi)^{d} }\,\, \ln \left( \frac{ p^{4} + (m^{2}+ N^{2})p^{2} + (m^{2}N^{2} + \lambda^{4}) }{p^{2}+N^{2}}  \right)
-  \frac{3\lambda^{4}d}{4g^{2}}\;,
\label{vacfreeenergy}
\end{eqnarray}
with $r_{\pm}^{2}$ standing for minus the roots of the denominator of the gluon propagator \eqref{rgzgluonprop}, $N^{2} = M^{2} + \rho_{1}$, and $I(m^{2},r,T)$  given by \eqref{defi}. Explicitly, the roots are
\begin{eqnarray}
r_{\pm}^{2} = \frac{ (m^{2}+N^{2}) \pm \sqrt{ (m^{2}+N^{2})^{2} - 4(m^{2}N^{2}+\lambda^{4}) } }{2}\;.
\end{eqnarray}
The (central) condensate values were extracted from \cite{Cucchieri:2011ig}:
\begin{subequations}\label{latticec}
\begin{align}
N^2 = M^2+\rho_1 &= \unit{2.51}{\giga\electronvolt\squared} \;, \\
m^2 &= \unit{-1.92}{\giga\electronvolt\squared} \;, \\
\lambda^4 &= \unit{5.3}{\power{\giga\electronvolt}4} \;.
\end{align}
Once again the vacuum energy will be minimized with respect to the Polyakov loop expectation value $r$. For the analysis of thermodynamic quantities, only contributions coming from terms proportional to $I(m^{2},r,T)$ will be needed. Therefore, we will always consider the difference ${\cal E}_{v}(T) - {\cal E}_{v}(T=0)$. Since in the present  (RGZ) prescription the condensates are given by the zero temperature lattice results \eqref{latticec} instead of satisfying gap equations, the divergent contributions to the free energy are subtracted, and no specific choice of renormalization scheme is needed. Furthermore, explicit dependence on the coupling constant seems to drop out of the one-loop expression, such that no renormalization scale has to be chosen. Following the steps taken in Section \ref{Tonafhankelijk}, we find a second order phase transition at the temperature:
\end{subequations}
\begin{equation}
T_\text{crit} = \unit{0.25}{\giga\electronvolt} \;,
\end{equation}
which is not that far from the value of the $SU(2)$ deconfinement temperature  found on the lattice: $T_c\approx\unit{0.295}{\giga\electronvolt}$, as
quoted in \cite{Cucchieri:2007ta,Fingberg:1992ju}.

In future work, it would in particular be interesting to find out whether
---upon using the RGZ formalism--- the Gribov mass and/or RGZ condensates
directly feel the deconfinement transition, similar to the cusp we
discovered in the Gribov parameter following the exploratory restricted
analysis of this paper. This might also allow to shed further light on the
ongoing discussion of whether the deconfinement transition should be felt at
the level of the correlation functions, in particular the electric screening
mass associated with the longitudinal gluon propagator \cite{Cucchieri:2012nx,Cucchieri:2012nxb,Cucchieri:2012nxc,Cucchieri:2012nxd}.

\begin{figure}[h]
\begin{center}
\includegraphics[width=.45\textwidth]{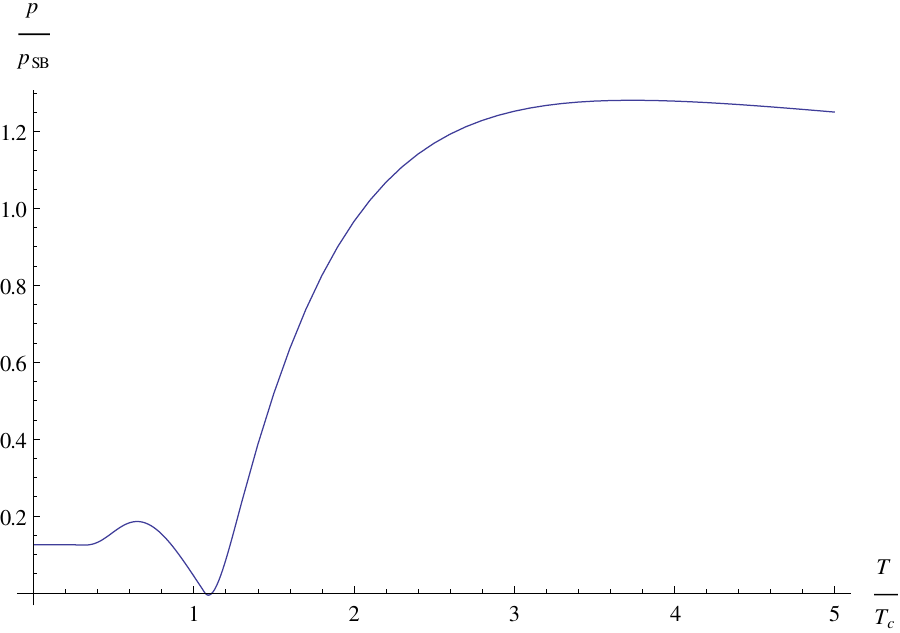} \hspace{10mm}
\includegraphics[width=.45\textwidth]{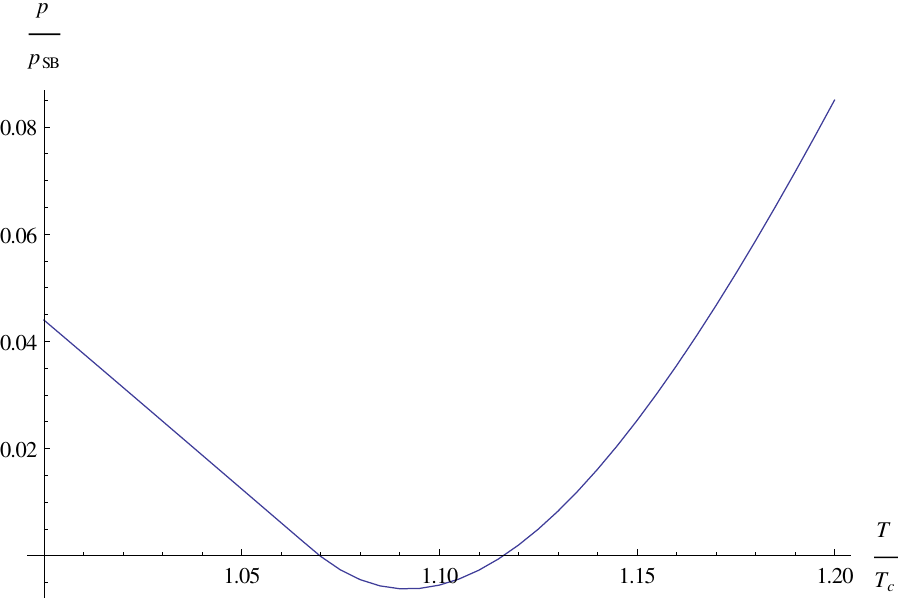}
\end{center}
\caption{Right and left plots refer to the RGZ pressure in terms of $T/T_{c}$ and in units of $T^{4}$. In the left plot, a wide temperature range of is shown. In the right plot, a zoom is made for temperatures around $1.10\; T_{c}$ to show the existence of negative pressure.
}
\label{rgzP}
\end{figure}

\begin{figure}[h]
\begin{center}
\includegraphics[width=.45\textwidth]{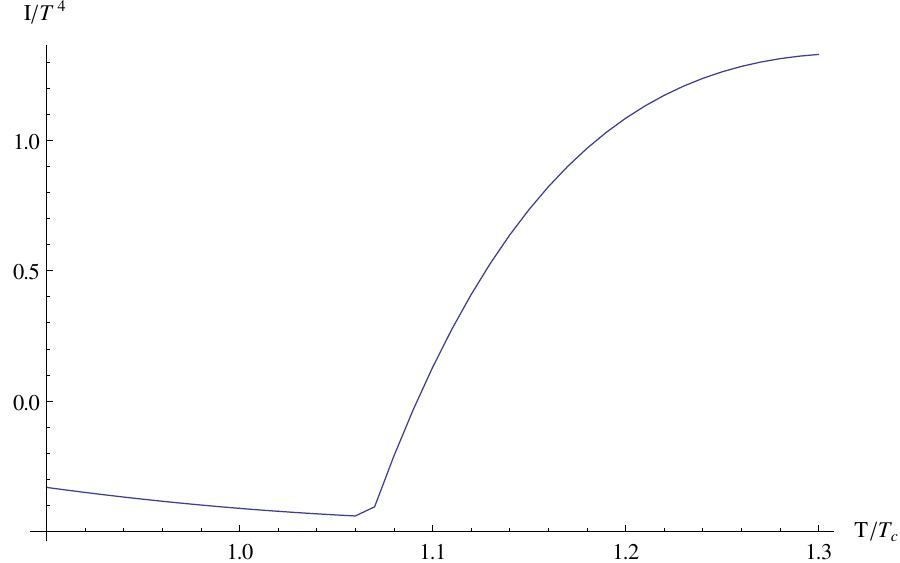}
\end{center}
\caption{The RGZ interaction measure $I/T^{4}$ in units $T/T_{c}$.}
\label{rgzTraceAnom}
\end{figure}

Let us also consider the pressure and interaction measure once more. The results are shown in \figurename\ \ref{rgzP} and \figurename\ \ref{rgzTraceAnom}, respectively. The oscillating behaviour at low temperature persists at leading order, while a small region of negative pressure is still present --- see the right plot of \figurename\ \ref{rgzP}. These findings are similar to \cite{Fukushima:2012qa} (low temperature results are not shown there), where two sets of finite temperature RGZ fits to the $SU(3)$ lattice data were used \cite{Aouane:2011fv,Aouane:2012bk}, in contrast with our usage of zero temperature $SU(2)$ data. In any case, a more involved analysis of the RGZ finite temperature dynamics is needed to make firmer statements. As already mentioned before, there is also the possibility that important low temperature physics is missing, as for instance the proposal of \cite{Fukushima:2012qa} related to the possible effect of light electric glueballs near the deconfinement phase transition \cite{Ishii:2001zq,Hatta:2003ga}. Obviously, these effects are absent in the current treatment (or in most other treatments in fact).

\section{Summary}

In this paper we studied the Gribov--Zwanziger (GZ) action for $SU(2)$ gauge theories with
the Polyakov loop coupled to it via the background field formalism. Doing
so, we were able to compute in a simultaneous fashion the finite temperature
value of the Polyakov loop and Gribov mass to the leading one-loop
approximation. The latter dynamical scale enters the theory as a result of
the restriction of the domain of gauge field integration to avoid
(infinitesimally connected) Gribov copies. Our main result is that we found
clear evidence of a second order deconfinement phase transition at finite
temperature, an occurrence accompanied by a cusp in the Gribov mass, which
thus directly feels the transition. It is perhaps worthwhile to stress here
that at temperatures above $T_{c}$, the Gribov mass is nonzero, indicating
that the gluon propagator still violates positivity and as such it rather
describes a quasi- than a \textquotedblleft free\textquotedblright\
observable particle, see also \cite{Maas:2011se,Haas:2013hpa} for more on this.

We also presented the pressure and trace anomaly, indicating there is a
problem at temperatures around the critical value when using the original GZ
formulation. We ended with a first look at the changes a full-fledged
analysis with the Refined Gribov--Zwanziger (RGZ) formalism might afflict, given
that the latter provides an adequate description of zero temperature gauge
dynamics, in contrast to the GZ predictions. This will be studied further in
upcoming work. Note that, even not considering finite temperature corrections to the condensates in the RGZ formalism, the region of negative pressure is considerably smaller than the region found with the GZ formalism.

A further result of the present paper, which is interesting from the
methodological point of view, is that it shows explicitly that
finite-temperature computations (such as the computation of the vacuum expectation value of the
Polyakov loop) are very suitable to be analyzed using analytical Casimir-like techniques. The interesting issue of Casimir-style computations at finite
temperatures is that, although they can be more involved, they provide one
with easy tools to analyze the high and low temperature limits as well as
the small mass limit. Moreover, within the Casimir framework, the
regularization procedures are often quite transparent. Indeed, in the
present paper, we have shown that the computation of the vacuum expectation value of the Polyakov
loop is very similar to the computation of the Casimir energy between two
plates. We believe that this point of view can be useful in different
contexts as well.

\section*{Acknowledgements}
We wish to thank P.~Salgado-Rebolledo for helpful discussions and comments. This work has been funded by the Fondecyt grants 1120352. The Centro de
Estudios Cientificos (CECs) is funded by the Chilean Government through the Centers of Excellence Base Financing Program of Conicyt. I.~F.~J.~is grateful for a PDSE scholarship from CAPES (Coordena{\c{c}}{\~{a}}o de Aperfei{\c{c}}oamento de Pessoal de N{\'{\i}}vel Superior, Brasil). P.~P. was partially supported from Fondecyt grant 1140155 and also thanks the Faculty of Mathematics and Physics of Charles University in Prague, Czech Republic for the kind hospitality at the final stage of this work.

\end{document}